\documentstyle[aps,epsfig,preprint]{revtex}

\begin{document}
\tighten
\draft
\preprint{
\vbox{
\hbox{March 1997}
\hbox{DOE/ER/40762-117}
\hbox{U.MD. PP\# 97-100}
}}

\title{SPIN-DEPENDENT TWIST--FOUR MATRIX ELEMENTS \\
	FROM $g_1$ DATA IN THE RESONANCE REGION}

\author{Xiangdong Ji and W. Melnitchouk}
\address{Department of Physics,
	University of Maryland,
	College Park, MD 20742}

\maketitle

\begin{abstract}
Matrix elements of spin-dependent twist-four operators are extracted 
from recent data on the spin-dependent $g_1$ structure function of the
proton and deuteron in the resonance region.
We emphasize the need to include the elastic contributions to the
first moments of the structure functions at $Q^2 < 2$ GeV$^2$.
The coefficients of the $1/Q^2$ corrections to the Ellis-Jaffe
sum rules are found to be $0.04 \pm 0.02$ and $0.03 \pm 0.04$
GeV$^2$ for the proton and neutron, respectively.
\end{abstract}


There has been much activity in recent years surrounding various
deep-inelastic spin sum rules, such as the Bjorken \cite{BJ} and
Ellis-Jaffe sum rules \cite{EJ}, which test our understanding of
the spin structure of the nucleon, as well as our ability to calculate
higher-order radiative corrections in QCD.
An intriguing issue in the study of these sum rules at moderate values
of the momentum transfer squared, $Q^2$ (say $Q^2 \sim 0.5$ to 3 GeV$^2$),
is that of higher-twist corrections.
In this paper, we extract the twist-four matrix elements from recent
data taken by the E143 Collaboration at SLAC \cite{E143} on the proton
and deuteron $g_1$ structure functions, and compare these with some
recent theoretical estimates.

Higher-twist corrections to the polarized deep-inelastic sum rules 
were first studied by Shuryak and Vainshtein \cite{SV}.
The coefficient functions were recalculated in Ref.\cite{JU}
and confirmed in Refs.\cite{BBK} and \cite{FRANKFURT}.
There is a long list of calculations and estimates of the
non-perturbative higher-twist matrix elements in the literature
\cite{JU,BBK,JJ,STRATMANN,SONG,SUM,LATTICE}. 
The interplay between the perturbation series at high
orders and higher-twist matrix elements was first discussed by Mueller 
in Ref.\cite{MUELLER}. 
The question concerns the precise definition of higher-twist corrections, 
as these are related to the procedure of how to regularize an asymptotic 
perturbation series. 
A concrete proposal of separating the perturbative and non-perturbative
contributions was suggested in Ref.\cite{RENORM}.

Very recently, the E143 Collaboration \cite{E143} published the first
data on the first moments of the $g_1$ structure functions of the
proton and neutron (the latter being extracted from deuterium data)
in the resonance region (at $Q^2 = 0.5$ and 1.2 GeV$^2$).
According to the phenomenon of parton--hadron duality, these data have 
a direct bearing on the size of the higher-twist matrix elements. 
{}From the successful phenomenology of the QCD sum rule method \cite{SVZ},
power corrections in an operator product expansion (OPE) have direct
control over the structure of the low-lying resonances.
In fact, much work has appeared in the literature on the calculation
of low-energy hadronic observables in terms of vacuum condensates.
Of course, the parton--hadron duality also allows determination of
the same condensates from the resonance masses and widths, if these
are known to sufficient accuracy.
In our case, we shall employ the latter approach to duality: namely,
resonances (including the nucleon elastic contribution) fix the higher-twist
matrix elements.
In the case of unpolarized deep-inelastic scattering, a similar analysis
was performed in Ref.\cite{JU95}.
Here we shall use the newly measured data from Ref.\cite{E143} to
extract the spin-dependent twist-four matrix elements.

The first moment of the $g_1$ structure function of the nucleon is
defined as:
\begin{equation}
\Gamma^N(Q^2)
= \int_0^1 dx\ g_1^N(x,Q^2)\ ,
\end{equation}
where $N = p$ or $n$, and the upper limit includes the nucleon elastic
contribution.
The inclusion of the elastic component is critical if one wishes to use the
OPE to study the evolution of the sum rule in the moderate $Q^2$ region
\cite{JI93}.
Note that the application of the OPE requires a product of two currents,
which arises in deep-inelastic scattering only after summing over all 
final (including elastic) hadronic states.
For unpolarized scattering, this point was emphasized some time 
ago by J. Ellis \cite{ELLIS}.
In Ref.\cite{BI}, attempts were made to extract the twist-four matrix
elements from the first moment of $g_1$, without inclusion of the 
elastic contribution.
In light of the logic behind the OPE, this procedure is clearly incorrect. 
Furthermore, the OPE is known to break down at low $Q^2$
($Q^2 \alt 0.5$ GeV$^2$); hence the Drell-Hearn-Gerasimov sum rule \cite{DHG}
at $Q^2=0$ has no obvious bearing on the size of the twist-four contributions
\cite{JU}.

In Figs.1 and 2 we show the first moments $\Gamma^p(Q^2)$ and 
$\Gamma^n(Q^2)$, respectively, at three different $Q^2$ values
($Q^2 \alt 3$ GeV$^2$) (the data are from Refs.\cite{E143,E143G1P,E142}).
The data, shown by the open circles, contain contributions from 
the measured $x$-regions covered in the experiments, together with
theoretical extrapolations into the unmeasured small-$x$ region,
and also from elastic scattering at $x=1$.
Note that the neutron points are obtained from the deuteron and proton
data assuming a small ($\sim 5\%$) D-state admixture in the deuteron.
Compared with the proton data, the neutron results are much
poorer, and could be significantly improved in future measurements
at Jefferson Lab \cite{MEZIANI}.
One should note that the contributions from the $x$-region covered by
the recent E143 analysis \cite{E143} are not entirely model independent.
A number of assumptions have been made regarding the contributions of
the $g_2$ structure function, the $\Delta$ resonance, etc. \cite{E143}.
For the present analysis, however, we simply adopt the data as published
in Ref.\cite{E143}, keeping in mind that future experiments may help to
clarify some of these assumptions.

For the small-$x$ extrapolation, we use a recent parameterization from 
Ref.\cite{BFR} fitted to the global polarized deep-inelastic scattering
data, which respects the Bjorken sum rule ($g_A = 1.257$ and 
$\alpha_s^{NLO}(M_Z^2) = 0.117 \pm 0.005$).
The parameterization has been constructed at a scale $Q_0^2=1$~GeV$^2$,
and evolution to different $Q^2$ values has been done including the 
complete NLO corrections \cite{WM}.
At such small $Q^2$ values one might question the role of higher-twist
contributions in any small-$x$ extrapolation.
However, previous experience from unpolarized deep-inelastic data tells
us that higher-twist effects at small $x$ tend to be rather small
\cite{MSTVV}.
It is difficult to assign an error to the theoretical small-$x$
extrapolation, as this is still a somewhat controversial, but 
interesting, subject which is currently under active study 
\cite{HERA,ABFR,BL,CR,BADELEK}.
The data shown by the squares in Figs.1 and 2 include only the
inelastic contributions, as discussed above.

The dotted curves in Figs.1 and 2 represent the elastic contributions
to the $\Gamma^N$ moments calculated in terms of nucleon form factors
\cite{JU}:
\begin{eqnarray}
\Gamma_{el}^{N}(Q^2)
&=& {1 \over 2} F_1^N(Q^2) \left( F_1^N(Q^2) + F_2^N(Q^2) \right),
\end{eqnarray}
where we have used the parameterization of $F_{1,2}^N(Q^2)$ from 
the fit of Ref.\cite{MMD}.
For the proton, the elastic contribution is negligible at $Q^2 > 3$ GeV$^2$,
and is about 10\% of the inelastic at $Q^2 = 2$ GeV$^2$.
At $Q^2 = 1$ GeV$^2$ it is as important as the inelastic component,
and below 0.5 GeV$^2$ the elastic contribution becomes dominant.
For the neutron, the elastic contribution peaks around $Q^2=0.5$ GeV$^2$,
and becomes quite small above $\approx 1.5$ GeV$^2$ and below
$\approx 0.1$ GeV$^2$.

According to the OPE, in the limit of large $Q^2 \gg \Lambda_{\rm QCD}^2$, 
$\Gamma^N(Q^2)$ can be calculated via the twist expansion:
\begin{equation}
\label{twist}
\Gamma^N(Q^2) = \sum_{\tau=2,4,\cdots} {\mu_\tau^N(Q^2) \over Q^{\tau-2}}\ ,
\end{equation}
where $\mu_\tau^N$ is related to nucleon matrix elements of operators
of twist $\leq \tau$.

{}From the total $\Gamma^N(Q^2)$ one can obtain the higher-twist
component by subtracting the twist-two contribution, $\mu_2^N$,
which can be written as a series expansion in $\alpha_s$:\ \ 
$\mu_2^N(Q^2) = \sum_n C_n^N \alpha_s(Q^2)$.
It is suspected that the coefficients $C_n^N$ grow like $n!$ as 
$n \rightarrow \infty$, and therefore, strictly speaking, $\mu_2^N$
is not a well-defined quantity \cite{MUELLER}.
The uncertainty in regularizing the divergent series is closely related 
to the precise definition of the higher-twist contributions.
In this paper we maximally utilize the available perturbative calculations
up to ${\cal O}(\alpha_s^3)$ \cite{LARIN}, and define $\mu_2^N$ up to 
this order as the entire twist-two contribution.
We will return to this point later.
For three quark flavors, the three-loop result for the twist-two component
of the proton and neutron first moments is given by \cite{LARIN}:
\begin{eqnarray}
\label{mu2}
\mu_2^{p(n)}(Q^2)
&=& \left[ 1 - \left( {\alpha_s \over \pi} \right)
	     - 3.5833 \left( {\alpha_s \over \pi} \right)^2	
	     - 20.2153 \left( {\alpha_s \over \pi} \right)^3
    \right]
\left( \pm {1 \over 12} g_A\ +\ {1 \over 36} a_8 \right)	\nonumber\\
&+& \left[ 1 - 0.3333 \left( {\alpha_s \over \pi} \right)
	     - 0.54959 \left( {\alpha_s \over \pi} \right)^2
	     - 4.44725 \left( {\alpha_s \over \pi} \right)^3
    \right]
{1 \over 9} \Sigma_{\infty}\ ,
\end{eqnarray}
where the $\pm$ refers to $p$ or $n$.
The leading-twist component in (\ref{mu2}) is given in terms of the
triplet and octet axial charges, $g_A$ and $a_8$, respectively, 
and the quantity $\Sigma_{\infty}$, defined as the renormalization 
group invariant nucleon matrix element of the singlet axial current 
\cite{LARIN}, $\Sigma_{\infty} \equiv \Sigma(Q^2=\infty)$.
With the above value of $g_A$, the leading-twist contribution (\ref{mu2})
is calculated with the values $a_8 = 0.579 \pm 0.025$ \cite{CR},
extracted from weak hyperon decays, and
$\Sigma_{\infty} \approx 0.15 \pm 0.12$
obtained from the global fit of Ref.\cite{BFR}.
The results for the proton and neutron are shown in Figs.1 and 2
by the solid curves, where the band range reflects the combined error
in $\Sigma_{\infty}$ and $\alpha_s$.
For the latter we have taken $\alpha_s^{NLO}(1$ GeV$^2) = 0.45 \pm 0.05$
\cite{SCHMELLING}.
Subtracting from the open circles in Figs.1 and 2 the leading-twist
component, one obtains the pure higher-twist contribution, denoted
by the full circles.
Note that the increased error bars on the higher-twist points simply
reflect the uncertainty in $\mu_2^N(Q^2)$.
Thus the higher-twist contribution and the uncertainty in
$\alpha_s$ are correlated at intermediate values of $Q^2$
($0.5 \alt Q^2 \alt 2$ GeV$^2$).
The higher-twist contribution can thus be reliably extracted only
when $\alpha_s$ is more accurately determined from other sources.

Turning now to the higher-twist contributions to $\Gamma^N(Q^2)$,
the $1/Q^2$ term in Eq.(\ref{twist}) is a sum of three terms:
\begin{eqnarray}
\label{mu4}
\mu_4^N(Q^2)
&=& {1 \over 9} M^2
\left( a_2^N(Q^2) + 4 d_2^N(Q^2) - 4 f_2^N(Q^2) \right).
\end{eqnarray}
The $a_2^N$ component, being given by the second moment of the twist-two
part of the polarized $g_1^N$ structure function,
\begin{eqnarray}
\label{a2N}
a_2^N(Q^2)
&=& 2 \int_0^1 dx\ x^2\ g_1^N(x,Q^2)\ ,
\end{eqnarray}
arises from the target mass correction \cite{JU}, and can be evaluated
straightforwardly from global parameterizations of the leading-twist
part of $g_1^N(x,Q^2)$.
The twist-3 correction in Eq.(\ref{mu4}) can be extracted from the 
leading-twist contributions to the following moment of the $g_1^N$
and $g_2^N$ structure functions:
\begin{eqnarray} 
\label{d2N}
d_2^N(Q^2)
&=& \int_0^1 dx\ x^2\ \left( 2 g_1^N(x,Q^2)\ +\ 3 g_2^N(x,Q^2) \right).
\end{eqnarray}
Recently the $d_2$ coefficients of the proton and deuteron have been
determined in Ref.\cite{E143D2} at an average value of $Q^2 = 5$ GeV$^2$, 
with the results:
$d_2^p = 0.0054 \pm 0.0050$, and
$d_2^D = 0.0039 \pm 0.0092$.
To a good approximation the neutron $d_2^n$ can be obtained from the
relation:\ 
$d_2^n = 2 d_2^D / (1-3/2~\omega_D) - d_2^p$,\ 
where $\omega_D$ is the deuteron D-state probability
(more sophisticated treatments \cite{NUCL} which account for
binding and Fermi motion effects are small in comparison with
the present error bars).
With $\omega_D = 5\%$ this gives
$d_2^n = 0.0030 \pm 0.020$ at $Q^2 = 5$ GeV$^2$.
Without any D-state correction \cite{MSS} the value would be 
around 25\% smaller.
To obtain $d_2$ at a different $Q^2$ one can use the leading logarithmic
evolution as computed in Refs.\cite{SV,KUYK}.
In principle, the anomalous dimensions for the singlet and non-singlet
components differ, so that one would require separate knowledge of the
singlet and non-singlet matrix elements to perform the evolution.
In practice, however, for 3 flavors the one-loop singlet and non-singlet 
anomalous dimensions turn out to be very similar \cite{SV,KUYK}, and at
the present level of accuracy these differences can be safely ignored.

Lastly, the twist-four contribution, $f_2^N$, to $\mu_4^N$ is defined
by the matrix element:
\begin{eqnarray}
2 f_2^N(Q^2) M^2 S^\mu
&=& \sum_f e_f^2 
\left< P,S \left| 
	g \overline{\psi}_f \widetilde{F}^{\mu\nu} \gamma_\nu \psi_f
\right| P, S \right>,
\end{eqnarray}
where $S^\mu$ is the nucleon spin vector, and the gluon field-strength
tensor is
$\widetilde{F}^{\mu\nu} = 1/2\ \epsilon^{\mu\nu\alpha\beta} F_{\alpha\beta}$,
with $\epsilon^{0123} = +1$, and $g$ appears in the covariant derivative
as $D^\mu = \partial^\mu + i g A^\mu$.

In the remainder of this paper, our focus will be on the extraction of
this matrix element from the data in Figs.1 and 2.
To achieve this, one must first subtract the leading-twist, target mass,
and the twist-3 contributions from the $\Gamma^N(Q^2)$ data, and define:
\begin{eqnarray}
\label{DelGam}
\Delta\Gamma^N(Q^2)
&\equiv& \Gamma^N(Q^2) - \mu_2^N(Q^2) 
 - {1 \over 9} {M^2 \over Q^2} \left( a_2^N(Q^2) + 4 d_2^N(Q^2) \right).
\end{eqnarray}
In Figs.3 and 4 the extracted $\Delta\Gamma^N(Q^2)$ values for the
proton and neutron are shown, respectively, as a function of $1/Q^2$.
Theoretically, the data in Figs.3 and 4 represent contributions from
$f_2^N(Q^2)$ as well as from $\tau=6$ and higher twists:
\begin{eqnarray}
\label{residue}
\Delta\Gamma^N(Q^2)
&=& - {4 \over 9 } { M^2 \over Q^2 } f_2^N(Q^2)\ 
 +\ \sum_{\tau=6,8\cdots} { \mu_\tau^N \over Q^{\tau - 2} }\ .
\end{eqnarray}
The twist expansion is believed, however, to be controlled by a scale
related to the average transverse momentum of quarks in the nucleon
\cite{GP}, typically of the order 0.4--0.5 GeV \cite{JU,GP}.
Therefore one can reasonably expect that the role of $\tau \geq 6$
effects should not be significant for $Q^2 > 1$ GeV$^2$, and not 
overwhelming for $Q^2 \agt 0.5$ GeV$^2$.

Finally, the matrix elements $f_2^N$ can be extracted from the
data points in Figs.3 and 4 at $Q^2 > 1$ GeV$^2$ by neglecting
the higher-twist terms in Eq.(\ref{residue}).
The $Q^2$ evolution of $f_2^N(Q^2)$ is also taken into account
at leading logarithmic order \cite{SV,KUYK}.
This logarithmic $Q^2$ dependence results in the slight deviations
in the curves in Figs.3 and 4 from linearity.
At a scale of $Q^2 = 1$ GeV$^2$ we find:
\begin{eqnarray}
\label{f2p}
f_2^p &=& -0.10 \pm 0.05\ ,  \\
\label{f2n}
f_2^n &=& -0.07 \pm 0.08\ .
\end{eqnarray}
To determine the effects of the higher-order terms in the coefficient
functions of the twist-two contributions, we have also performed the
extraction by neglecting the ${\cal O}(\alpha_s^3)$ terms in 
Eq.(\ref{mu2}).
The central value of the twist-four matrix element for the proton
is then reduced from $-0.10$ to $-0.08$, while for the neutron it
increases slightly, from $-0.07$ to $-0.08$.
Including the $a_2^N$ and $d_2^N$ contributions, one can finally 
determine the $1/Q^2$ correction to the Ellis-Jaffe sum rules:
\begin{eqnarray}
\mu_4^p &=& (0.04 \pm 0.02)\ {\rm GeV}^2\ , \\
\mu_4^n &=& (0.03 \pm 0.04)\ {\rm GeV}^2\ , 
\end{eqnarray}
at the scale $Q^2 = 1$ GeV$^2$.

The central values in Eqs.(\ref{f2p}) and (\ref{f2n}) seem to suggest
that the isoscalar combination of the twist-four matrix elements is much
larger than the isovector combination.
One might suspect therefore that the singlet twist-two contribution to 
the sum rule obtained from the global fit \cite{BFR} is too small.
To investigate the effect that a larger value of $\Sigma_{\infty}$
would have on the twist-four matrix elements, we have reanalyzed
the data using $\Sigma_{\infty} \approx 0.3$ as the central value
\cite{BADELEK}.
The effect is a reduction of $f_2^p$ to $\approx -0.05$, and 
$f_2^n$ to $\approx 0.0$, which would then lead to similar 
values for both the isotriplet and isosinglet combinations.
Of course these values are still consistent with the results in 
Eqs.(\ref{f2p}) and (\ref{f2n}) within the errors.
In principle, one could eliminate the dependence on $\Sigma_{\infty}$
by considering only the isovector combination $\Gamma^p-\Gamma^n$,
thereby reducing significantly the uncertainty in the isovector 
twist-four matrix element.
Unfortunately, the error associated with the neutron data is largely
experimental, so that the final proton---neutron moment would have an
error which is as large as that for the neutron points in Fig.4.

The values determined in Eqs.(\ref{f2p}) and (\ref{f2n}) can be
compared with several model calculations of the twist-four matrix 
elements in the literature.
The first estimates of $f_2^N$ were made using QCD sum rules.
The result from Ref.\cite{BBK} is
$f_2^p = 0.050 \pm 0.034$ and $f_2^n = -0.018 \pm 0.017$, 
while that from Ref.\cite{SUM} is
$f_2^p = 0.037 \pm 0.006$ and $f_2^n = 0.013 \pm 0.006$.
Alternative estimates of the $\tau=4$ matrix elements were made
using the MIT bag model \cite{JU,JIPOL}.
The result there, evolved from the bag scale up to $Q^2 \sim 1$ GeV$^2$,
was found to be $f_2^p = -0.028$ and $f_2^n = 0$.

The results obtained in this work will be improved as more 
experimental information becomes available in future.
The error on the data points in Figs.3 and 4 come mainly from
the uncertainty associated with the value of $\alpha_s$, and the
singlet axial charge $\Sigma_{\infty}$, when subtracting the
twist-two contribution from the total $\Gamma^N(Q^2)$.
Therefore better knowledge of the twist-two part of the structure
function at higher $Q^2$, and a more accurate determination of 
$\alpha_s$ from other experiments, will be valuable in pinning
down the higher-twists at low $Q^2$. 
Certainly more data points are needed in order to establish a clearer
trend of the $Q^2$ dependence at moderate values of $Q^2$
($Q^2 \sim 0.5 - 3.0$ GeV$^2$).
In particular, the neutron data points are irregular and should 
be confirmed in subsequent experiments.
In this respect, future experiments at Jefferson Lab and other
facilities can contribute much to our present understanding of
the twist-four matrix elements of the nucleon.

\acknowledgements

This work was supported by the DOE grant DE-FG02-93ER-40762.


\begin{figure}
\label{fig1}
\epsfig{figure=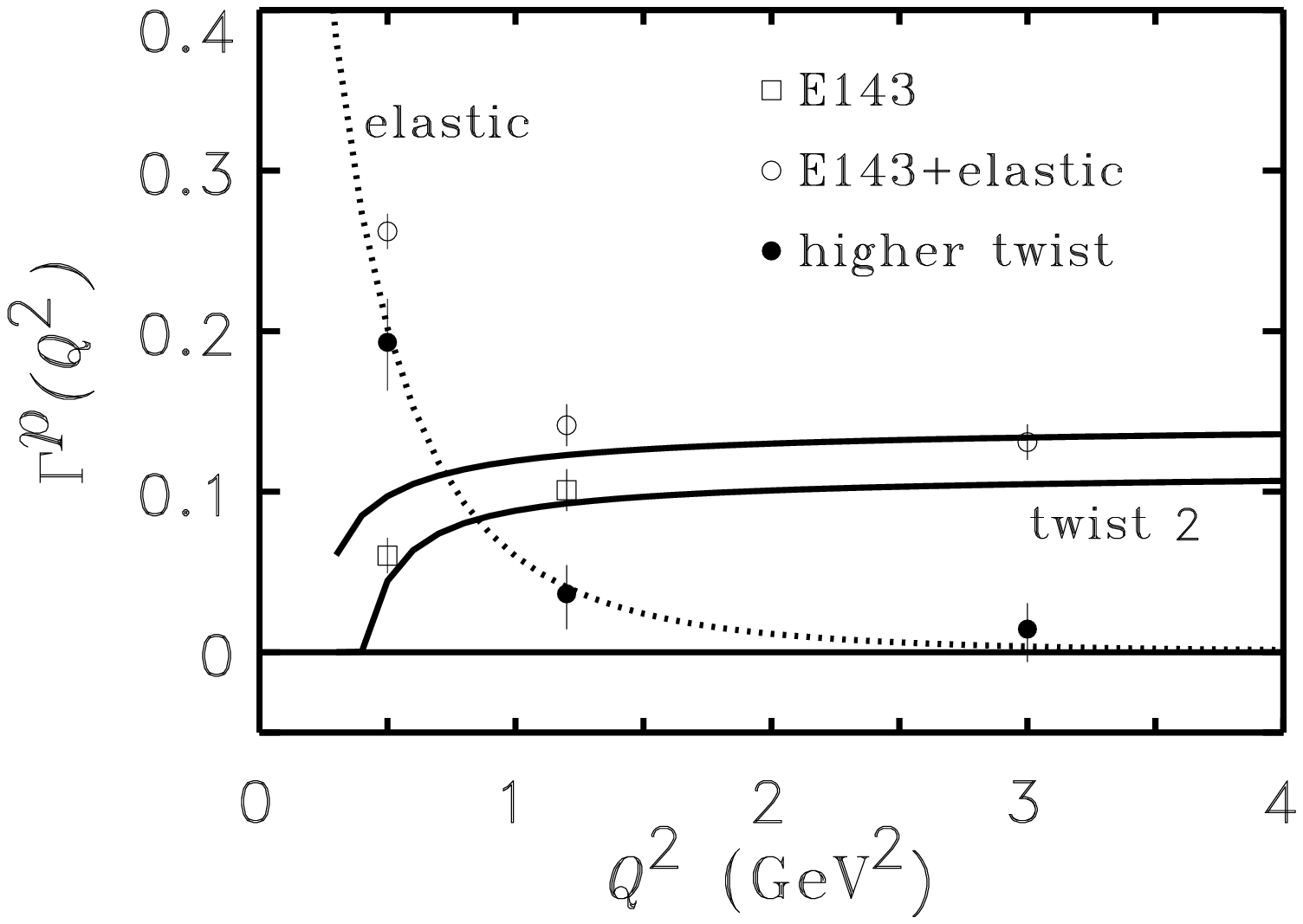,height=9cm}
\caption{$Q^2$ dependence of $\Gamma^p(Q^2)$.
	The solid curves represent the twist-two part of $\Gamma^p(Q^2)$,
	while the dotted is the elastic component,
	as parameterized in Ref.\protect\cite{MMD}.
	The squares denote the inelastic contribution extracted from the
	E143 experiment \protect\cite{E143}, the open circles include also
	the elastic piece, while the full circles are the pure higher-twist
	contributions.}
\end{figure}

\begin{figure}
\label{fig2}
\epsfig{figure=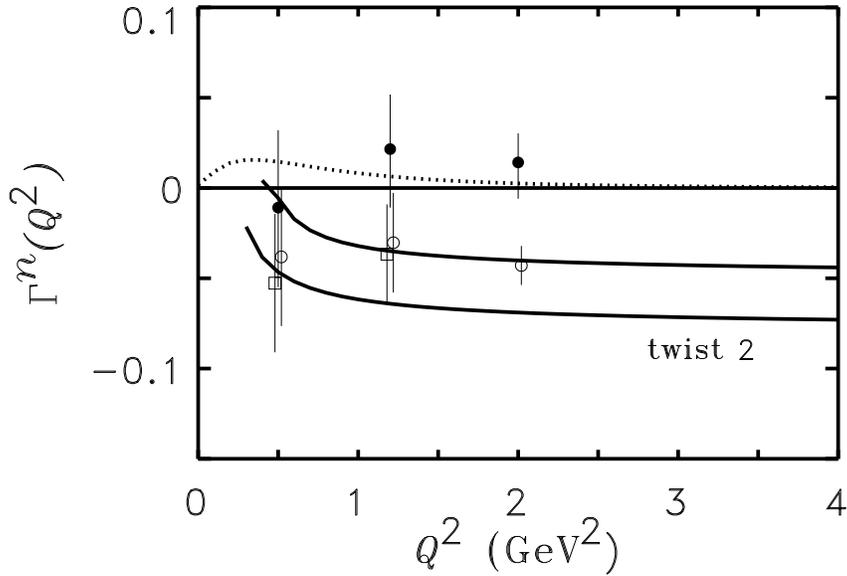,height=9cm}
\caption{Same as in Fig.1, but for the neutron.}
\end{figure}

\begin{figure}
\label{fig3}
\epsfig{figure=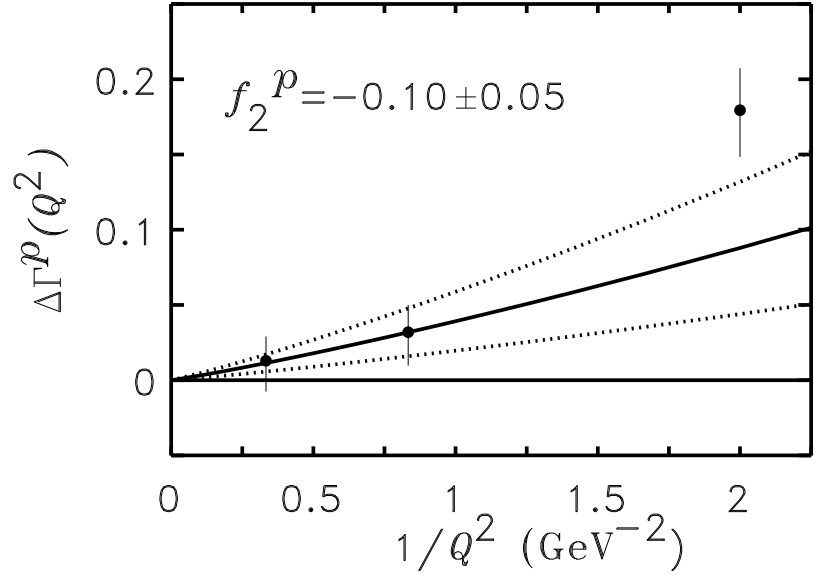,height=9cm}
\caption{The $1/Q^2$ dependence of the higher-twist contribution
	$\Delta\Gamma^p(Q^2)$ defined in Eqs.(\protect\ref{DelGam})
	and (\protect\ref{residue}).
	The curves correspond to $f_2^p = -0.10 \pm 0.05$ at
	$Q^2 = 1$ GeV$^2$ (the solid represents the central value,
	while the dotted curves indicate the error range on $f_2^p$).}
\end{figure}

\begin{figure}
\label{fig4}
\epsfig{figure=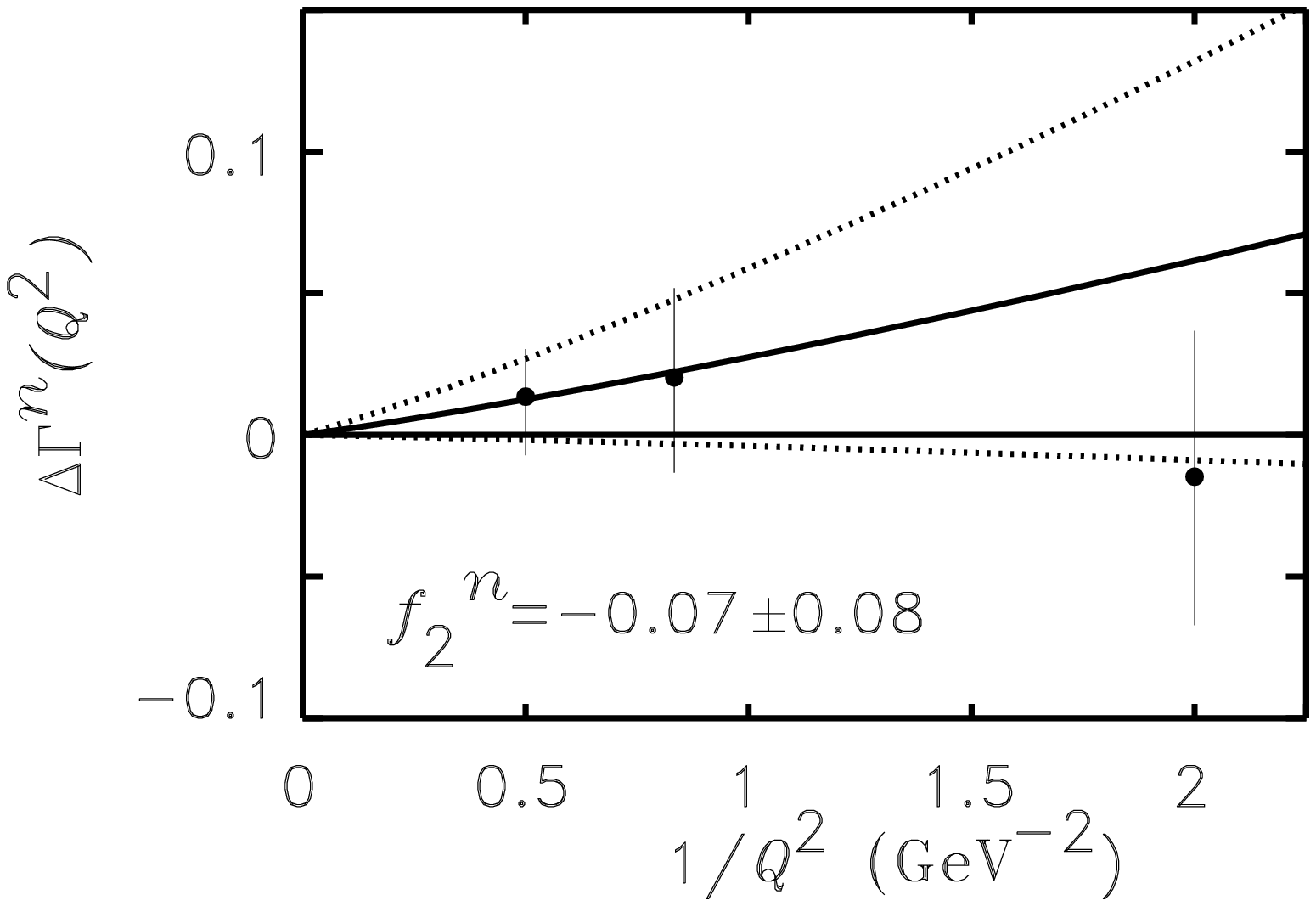,height=9cm}
\caption{Same as in Fig.3 but for the neutron.
	The curves correspond to $f_2^n = -0.07 \pm 0.08$
	at $Q^2 = 1$ GeV$^2$.}
\end{figure}


\begin{references}

\bibitem{BJ}
J. D. Bjorken,
Phys. Rev. 148 (1966) 1467.

\bibitem{EJ}
J. Ellis and R. L. Jaffe,
Phys. Rev. D 9 (1974) 1444;
Erratum-ibid. D 10 (1974) 1669.

\bibitem{E143}
E143 Collaboration, K. Abe et al.,
SLAC preprint SLAC-PUB-7242, Sep 1996,
hep-ex/9701004.

\bibitem{SV}
E. V. Shuryak and A. I. Vainshtein,
Nucl. Phys. B201 (1982) 141.

\bibitem{JU}
X. Ji and P. Unrau,
Phys. Lett. B 333 (1994) 228.

\bibitem{BBK}
I. I. Balitsky, V. M. Braun and A. V. Kolisnichenko,
JETP Lett. 50 (1989) 61;
Phys. Lett. B 242 (1990) 245;
Erratum-ibid. B 318 (1993) 648.

\bibitem{FRANKFURT}
B. Ehrnsperger, L. Mankiewicz and A. Sch\"afer,
Phys. Lett. B 323 (1994) 439.

\bibitem{JJ}
R. L. Jaffe and X. Ji,
Phys. Rev. D 43 (1991) 724.

\bibitem{STRATMANN}
M. Stratmann,
Z. Phys. C 60 (1993) 763.

\bibitem{SONG}
X. Song and J. S. McCarthy,
Phys. Rev. D 49 (1994) 3169;
X. Song,
Phys. Rev. D 54 (1996) 1955.

\bibitem{SUM}
E. Stein, P. Gornicki, L. Mankiewicz, A. Sch\"afer and W. Greiner,
Phys. Lett. B 343 (1995) 369;	
%
E. Stein, P. Gornicki, L. Mankiewicz and A. Sch\"afer,
Phys. Lett. B 353 (1995) 107.		

\bibitem{LATTICE}
M. G\"ockeler, R. Horsley, E.M. Ilgenfritz, H. Perlt, P. Rakow,
G. Schierholz and A. Schiller,
Phys. Rev. D 53 (1996) 2317.

\bibitem{MUELLER}
A. Mueller,
Phys. Lett. B 308 (1993) 355.

\bibitem{RENORM}
X. Ji,
MIT-CTP-2437, Jun 1995,
hep-ph/9506216.

\bibitem{SVZ}
M. A. Shifman, A. I. Vainshtein and V. I. Zakharov,
Nucl. Phys. B147 (1979) 385; 447.

\bibitem{JU95}
X. Ji and P. Unrau,
Phys. Rev. D 52 (1995) 72.

\bibitem{JI93}
X. Ji,
Phys. Lett. B 309 (1993) 187.

\bibitem{ELLIS}
J. Ellis, 
talk presented at Neutrino 79, 
CERN preprint TH.2701-CERN (1979).

\bibitem{BI}
M. Anselmino, B. L. Ioffe and E. Leader,
Yad. Fiz. 49 (1989) 214;
%
V. D. Burkert and B. L. Ioffe,
Phys. Lett. B 296 (1992) 223;
J. Exp. Theor. Phys. 78 (1994) 619.

\bibitem{DHG}
For recent reviews see
D. Drechsel,
Prog. Part. Nucl. Phys. 34 (1995) 181;
%
S. D. Bass,
Bonn report TK-96-29, March 1997.

\bibitem{MMD}
P. Mergell, U.-G. Mei\ss ner and D. Drechsel,
Nucl. Phys. A596 (1996) 367.

\bibitem{E143G1P}
E143 Collaboration, K. Abe et al.,
Phys. Rev. Lett. 74 (1995) 346.

\bibitem{E142}
E142 Collaboration, P. L. Anthony et al.,
Phys. Rev. Lett. 71 (1993) 959.

\bibitem{MEZIANI}
Z. E. Meziani,
talk presented at Baryons '95, 
Proceedings of the 7th International Conference on the Structure of Baryons,
eds. B. F. Gibson et al.
(World Scientific, 1996).

\bibitem{BFR}
R. D. Ball, S. Forte and G. Ridolfi,
Phys. Lett. B 378 (1996) 255.

\bibitem{WM}
T. Weigl and W. Melnitchouk,
Nucl. Phys. B465 (1996) 267.

\bibitem{MSTVV}
A. Milsztajn, A. Staude, K.-M. Teichert, M. Virchaux and R. Voss,
Z. Phys. C 49 (1991) 527.

\bibitem{HERA}
Workshop on the Proceedings of Spin Physics at HERA,
Zeuthen, Aug.28-31, 1995,
eds. J. Bl\"umlein and W.-D. Nowak.

\bibitem{ABFR}
G. Altarelli, R. D. Ball, S. Forte and G. Ridolfi,
CERN-TH-96-345, Dec 1996,
hep-ph/9701289.

\bibitem{BL}
S. D. Bass and P. V. Landshoff,
Phys. Lett. B 336 (1994) 537.

\bibitem{CR}
F. E. Close and R. G. Roberts,
Phys. Lett. B 336 (1994) 257.

\bibitem{BADELEK}
B. Badelek,
talk given at Cracow International Symposium on Radiative Corrections
(CRAD 96), Cracow, Poland, August 1996,
hep-ph/9612274.

\bibitem{LARIN}
S. A. Larin, T. van Ritbergen and J. A. M. Vermaseren,
UM-TH-97-02, NIKHEF-97-011;
%
S. A. Larin,
Phys. Lett. B 334 (1994) 192.

\bibitem{SCHMELLING}
M. Schmelling,
Plenary talk given at the XXVIII International Conference
on High Energy Physics, Warsaw, July 25-31, 1996,
hep-ex/9701002.

\bibitem{E143D2}
E143 Collaboration, K. Abe et al.,
Phys. Rev. Lett. 76 (1996) 587.

\bibitem{NUCL}
W. Melnitchouk, G. Piller and A. W. Thomas,
Phys. Lett. B 346 (1995) 165;
%
G. Piller, W. Melnitchouk and A. W. Thomas,
Phys. Rev. C 54 (1996) 894;
%
S. A. Kulagin, W. Melnitchouk, G. Piller and W. Weise,
Phys. Rev. C 52 (1995) 932.

\bibitem{MSS}
L. Mankiewicz, E. Stein and A. Sch\"afer,
talk given at Workshop on the Prospects of Spin Physics,
Zeuthen, Germany, Aug 28-31, 1995,
hep-ph/9510418.

\bibitem{KUYK}
H. Kawamura, T. Uematsu, Y. Yasui and J. Kodaira,
Kyoto U. preprint KUCP-88, March 1996,
hep-ph/9603338.

\bibitem{GP}
H. Georgi and H. D. Politzer, 
Phys. Rev. D 14 (1976) 1829.

\bibitem{JIPOL}
X. Ji,
MIT preprint MIT-CTP-2468, hep-ph/9509288.

\end{references}
\end{document}